\documentstyle[twoside,fleqn,espcrc2]{article}
\input BoxedEPS
\SetRokickiEPSFSpecial  
\HideDisplacementBoxes

\newcommand{\Tr}{\mathop{\rm Tr}}  

\newcommand{\Max}{\mathop{\rm Max}\nolimits}
\newcommand{\tr}{\mathop{\rm tr}}
\newcommand{\half}{{\textstyle\frac12}}
\newcommand{\lr}[1]{\langle #1\rangle}
\let\epsilon\varepsilon

\newcommand{\AmS}{{\protect\the\textfont2
  A\kern-.1667em\lower.5ex\hbox{M}\kern-.125emS}}

\title{Center Vortices, Instantons, and Confinement\thanks{Talk presented
 by J.W. Negele}}
\author{O. Jahn,\address{Institute for Theoretical Physics III,
University of Erlangen-N\"urnberg,\\
Staudstrasse 7, 91058 Erlangen, Germany}
F. Lenz,$^{\rm a}$  J.W. Negele,\address{Center for Theoretical Physics, 
Laboratory for Nuclear Science, and Department of Physics, \\
Massachusetts Institute of Technology, 77 Massachusetts Ave., Cambridge,
MA 02139, USA}
and
M. Thies$^{\rm a}$}

\begin{document} 

\begin{abstract}\noindent
We study the relation between center vortices and instantons in lattice QCD.

\end{abstract}

\maketitle

\section{Motivation}

One approach to understanding the physical origin of confinement and
chiral symmetry breaking is to attempt to capture the essential physics
in a much smaller number of degrees of freedom by projecting from a
maximal abelian or maximal center gauge. For simplicity, we consider
$SU(2)$ and denote a link variable transformed by a gauge
transformation~$g$ by 
\begin{eqnarray}
U_\mu^g(x) &\equiv& g^\dag(x) U_\mu(x) g(x+\mu)\nonumber\\ 
&\equiv& \bigl(a_0^g +
i\sum_k a_k^g
\sigma_k\bigr)_{x,\mu}
\label{JW:eq:1}
\end{eqnarray}
with $\sum a_\mu^2 =1$. Maximal abelian gauge corresponds to calculating
\begin{eqnarray*}
&&\Max_{\{g\}} \sum_{\mu,\nu} \tr \bigl(U_\mu^g (x)\sigma_3
U_\mu^{g\dag}(x)\sigma_3\bigr)
 \\
&&\qquad = \Max_{\{g\}} \sum_{x,\mu}\bigl\{ 2\bigl[ (a_0^g)^2 +
(a_3^g)^2\bigr]_{x,\mu} -1 \bigr\}
\end{eqnarray*}
and the abelian projection is performed by truncating $a_1$ and $a_2$ and
normalizing to obtain a $U(1)$ matrix\cite{Wiese1}. Although results are
ambiguous because of Gribov ambiguities associated with local minima, the
projected results generally reproduce the proper string tension.

Maximal center gauge corresponds to calculating 
$$
\Max_{\{g\}} \sum_{x,\mu}  \bigl(\Tr U_\mu^g (x)\bigr)^2\!\!
 = \Max_{\{g\}} \sum_{x,\mu} 4  (a_0^2)_{x,\mu}  
$$
and center projection is performed by truncating $a_1$, $a_2$, and $a_3$
and normalizing the resulting $Z(2)$ variable to $\pm1$. The properties
of center projection have been discussed widely by Greensite and
collaborators\cite{Greensite2} and Reinhardt and
collaborators\cite{Reinhardt3}. The set of negative links defines a set
of center vortices, corresponding to closed lines in three
dimensions and closed surfaces in four dimensions. Vortices may also be
thought of as Dirac strings connecting monopoles. 

For subsequent reference, it is useful to visualize a vortex sheet
generated by a single negative link  in the $z$~direction from
$(0,0,0,0)$ to $(0,0,1,0)$. The vortex sheet is the surface of the cube
dual to this link, that is,  the unit cube in $(x,y,t)$ centered at the
origin and having $z=\half$. If one cuts this four-dimensional surface at
time $t=0$ and observes it in the three-dimensional $x,y,z$ space, it
corresponds to a square loop piercing the center of each of the
plaquettes with Wilson loop~$-1$ that share the negative-$z$ link. If two
adjacent links in the $z$~direction are~$-1$, the cube grows to a
rectangular parallelepiped, and in general surfaces can grow by stacking
the dual cubes together. Returning to our single cube dual to a $z$~link,
we note for example that the $x$-$t$ surface corresponds to a $y$-$z$
plaquette and thus corresponds to $B_x \sim F_{yz}$. In general, the
vortex sheet in the $ij$ plane corresponds
to $\epsilon_{ijk\ell}F_{k\ell}$. By suitably choosing the time direction
in euclidean space to either be in or perpendicular to the plane of the
sheet, we can think of the vortex sheets as being sheets of magnetic or
electric fields. Hence, a single smoothly bending vortex sheet corresponds
locally to either $\vec E$ or
$\vec B$, but can never generate $\vec E\cdot \vec B$. The simplest way
to generate lumps of $\vec E\cdot \vec B$, corresponding to the
topological charge density one expects from instantons, is to have an
intersection of two mutually perpendicular sheets. For example, a sheet
in the $x$-$y$ plane at fixed $(t_0, z_0)$ with $F_{zt}\sim E_z$
intersecting a sheet in the $z$-$t$ plane at fixed $(x_0, y_0)$ with
$F_{xy}\sim B_z$ yields a point of $\vec E\cdot \vec B$ at $(x_0, y_0,
z_0, t_0)$. If the vortices have finite thickness, the resulting lump of
$\vec E\cdot \vec B$ will have finite size. 

Lattice phenomenology provides several indications that the reduced
degrees of freedom subsequent to center projection still embody the
essence of confinement and chiral symmetry breaking. The contribution of
each vortex piercing a Wilson loop is $(-1)$, so that a random ensemble of
vortices produces confinement and lattice measurements yield the proper
string tension and scaling\cite{Greensite2,Reinhardt3}. There are several
indications that there are underlying physical structures (``thick
vortices'') corresponding to the ``thin vortices'' obtained by
projection. Calculations with twisted boundary conditions to prevent
decay to the vacuum yield stationary vortex solutions that are
exponentially localized in two dimensions and periodic in the
other two dimensions\cite{GM4}.  When the contribution of a lattice gluon
configuration to a Wilson loop
$W_n$ is labeled by the number~$n$  of projected vortices that would
pierce the loop, one observes that $\lr{W_2}/\lr{W_0} \to 1$ and 
 $\lr{W_1}/\lr{W_0} \to -1$ for large loops\cite{Greensite2}. This
suggests that there is a finite size object in the actual configuration
that, when enclosed by a large enough loop, behaves just like an
idealized ``thin vortex''. In addition, at finite temperature, the
vortices appear to become aligned in the time direction, so that they
pierce spatial loops yielding an area law. Finally, de~Forcrand and
D'Elia\cite{dfDE5} have studied the effect of removing center vortices.
For each link $U_\mu(x)$ one determines $Z_\mu(x)$ by center projection
and defines the modified distribution $U_\mu'(x) = Z_\mu(x) U_\mu(x)$,
which projects to the trivial configuration. Not only does one lose
confinement, but in addition $\lr{\bar\psi \psi}\to 0$ and the
topological charge $\lr Q \to 0$. That $\lr Q \to 0$ is understandable,
since $U$ must be far from~1 somewhere for nontrivial topology and
projection leads to the trivial sector. That $\lr{\bar\psi \psi}\to 0$
implies via  the Banks Cacher relation that the zero modes and thus all
the instantons and antiinstantons must also be removed.

\section{Center Projection of a Single Instanton}
To understand how the modified configuration removes instantons, we have
studied the center projection of a single instanton on a lattice.
Projection yields a cluster of parallel $-1$ links in the center of the
instanton in one common direction that we will call time. This breaks the
euclidean invariance of the instanton, and yields a vortex sheet that is
pure~$E$. The most relevant configurations are  a single dual cube, a
$2^3$ cube composed of 8 elementary cubes, and a $3^3$ cube composed of
27 elementary cubes. To see how the modified distribution cancels an
instanton, note that there must be a shift in group space of $\pi$ so
$U\sim e^{iE}\to e^{i(E+\pi)}$. Since
$$
Q^{\rm Lat} \sim \frac1{4\pi^2} \sum_{a,i} 2 E_i^a B_i^a
$$
if we assume that the full shift of $\pi$ 
occurs in a single plaquette for  an instanton of size~$\rho$ and assume a
spherical sheet of radius~$R$,
$$
Q^{\rm Lat} \sim \frac1{4\pi^2} \sum_{\rm vortex} 2(\pi \hat E_i^a) B_i^a
=\frac{2\rho^2R^2}{\rho^2 + R^2}
$$
so that $\delta Q=1$ when $R=\rho$. For a cube, $\delta Q= 1.203$ for
$R=0.813\,\rho$. Hence, if the size of the vortex is $\sim \rho$, it
should cancel~$Q$. 

The primary technical problem is local minima corresponding to Gribov
ambiguities. We start with a cooled instanton on lattices from $16^4$ to
$32^4$, minimize
$$
R_{\{g\}} \equiv \sum_{x,\mu} \bigl[1-\bigl(\half\tr g^\dag(x) U_\mu(x)
g(x+\mu)\bigr)^2\bigr]
$$
successively on each site by matrix diagonalization, and overrelax by
applying $g^\alpha$ with $\alpha\sim 1.7$ to~$1.8$. The local minimum
depends upon~$\alpha$. For example, a regular gauge instanton with
$\rho=4$ on a $16^4$ lattice yields no vortex for $\alpha=1.7$ and a $2^3$
vortex for $\alpha=1.8$. Similarly the initial gauge matters. A random
gauge transformation at $\alpha=1.7$ yields a $2^3$ vortex whereas the
regular gauge does not. Also, going first to Lorentz gauge can prevent
finding a vortex solution. By exploring a variety of gauges and initial
conditions, the $2^3$ vortex emerged as the minimum solution for this
case. 

\begin{figure}[tp]
\vspace*{-2mm}
$$
\mkern-10mu\BoxedEPSF{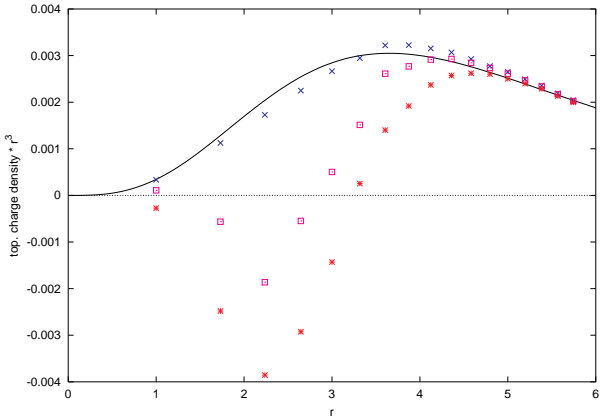 scaled 450}  
$$
\vspace*{-3pc}
\caption{Spherical average of topological charge density times $r^3$
versus radius $r$. The solid curve denotes a cooled instanton, the
crosses indicate the density after removal of a $4^3$ vortex, and the
squares and stars denote the density after one and two cooling steps,
respectively.}
\label{JNfig1}  
\vspace*{-.5cm} 
\end{figure}

As expected from the analytic estimate of $\delta Q^{\rm Lat}$, for a vortex
size~$R$ comparable to~$\rho$, center projection removes the instantons.
Figure~\ref{JNfig1} shows $r^3$ times the topological charge density as a
function of distance for a cooled instanton of size $\rho=4.7$, the
instanton with a vortex of size $R=4$ removed, and the density after one
and two cooling sweeps. Since $\delta Q$ begins localized on a shell, it
takes several cooling sweeps to see that the integral averages to zero.
After more cooling sweeps, the entire density relaxes to zero. A similar
calculation with size $R=2$ cools back to the original instanton with
$Q=1$. 

With hindsight, we believe this single-instanton example is an
unrepresentative special case. The problem is that the vortex sheet is a
pure $E$~field. Although when it has the proper area, it can combine with
the existing $B$~field of the instanton to cancel the original toplogical
charge, the vortex sheet by itself has no topological charge and thus
fails to represent the essential phyiscs of the original configuration. 

 As argued above we expect lumps of
$\vec E\cdot
\vec B$ to  arise from intersections of two vortices.
Whereas an isolated instanton will not have vortex sheets extending to
infinity, it would be natural for vortices to connect neighboring
instantons and antiinstantons.
 Thus,
the relevant case to examine is an ensemble of instantons and
antiinstantons, and to see whether projection produces intersections with
localized instanton-like lumps of $\vec E\cdot \vec B$. Were this to
occur, center vortices would succinctly represent the essential physics of
both confinement and chiral symmetry breaking.

\subsection*{Acknowledgments}

This work is supported in part by the U.S. Department of Energy (DOE)
under cooperative research agreement \#DE-FC02-94ER40818.

\end{document}